\newcommand{\be}{\begin{equation}}
\newcommand{\ee}{\end{equation}}
\newcommand{\bea}{\begin{eqnarray}}
\newcommand{\eea}{\end{eqnarray}}

\documentclass[aps,pral,twocolumn,groupedaddress]{revtex4}
\usepackage{graphicx}

\begin{document}

\title{Molecular Signatures in Hybrid Atomic/Molecular Bose-Einstein
Condensates}



\author{Chi-Yong Lin}
\affiliation{Department of Physics, National Dong Hwa
University Hua-Lien, Taiwan, R.O.C.}

\author{Paolo Tommasini, E.J.V. de Passos, M.S. Hussein,
A.F.R. de Toledo Piza}

\affiliation{
Instituto de F\'{\i}sica, Universidade de S\~ao Paulo, CP 66318
CEP 05389-970, S\~ao Paulo, \ SP, \ Brazil}

\author{E. Timmermans}

\affiliation{
T-4, Los Alamos National Laboratory Los Alamos, NM 87545, \ U.S.A.}

\date{\today}


\begin{abstract}

As it was proposed and recently verified experimentally, the
mechanism of Feshbach resonance in a condensate can create a
second condensate component of molecules that coexists with the
atomic condensate. In this work we investigate signatures of 
the presence of the molecular condensate through the equilibrium 
properties and collective excitations of the hybrid system 
of atoms and molecules, subjected to a trap, employing a 
time-dependent variational ansatz. We show that the shape 
of the condensate changes significantly by the presence 
of the molecules and that modes unique to this hybrid
system appear at observable frequencies.

\end{abstract}


\maketitle


The recent realization of Bose-Einstein condensation (BEC) in a
dilute gas of alkali atoms has open a new opportunity for the
theoretical and experimental investigation in quantum degenerate
many-body systems \cite{An95,DGPS99}. In contrast to superfluid
helium, these weakly interacting gases are much more amenable to
theoretical prediction and quantitative analysis. After the
observation of such condensates, a new generation of more complex
experiments has been done involving the production of multiple
species BEC. Such mixture may consist of different hyperfine states
of the same atom or different spin state of the same atomic species. In
the former case the composition of mixture remains fixed \cite{Ha98}
and in the latter case the internal conversion leads to important new
phenomena. It has been proposed that the interaction through the
Feshbach resonance that bring a binary atom system to an intermediate
state molecule recently observed in BEC\cite{In98},
create in the dilute atomic Bose-Einstein condensate (BEC) a second
molecular condensate component\cite{TTHK98,Dr98}. 
We should mention also  
the case of a dilute atomic BEC coupled to a molecular Bose gas by coherent
Raman transitions, which may lead to what is coined superchemistry
\cite{Wy00,HWDK00}. 

Recently, several groups have been working in order to detect a
signature of molecule formation in BEC  with Feshbach
resonance \cite{Wieman00,HO01}. 
Since these systems allow experimentalists to change the magnetic
filed (and consequently the detuning) one can probe these systems
in two different ways: 1 - adiabatic, where the detuning can be
changed slowly (waiting for the system to equilibrate) and 2 -
non-adiabatic, where a sudden change of the magnetic field is
performed (usually with a pulse). The non-adiabatic change leaving
the system out of equilibrium  will produce oscillations, among other
effects, in the number of atoms, and that was indeed observed \cite{Do02}.

In the adiabatic change the system is always at the equilibrium
configuration. In this paper we show that in this regime the shape of
the equilibrium condensate cloud of atoms and the behavior of the
excitation energies of the collective modes can be a signature of
the presence of molecules. It will be shown that the effects of the
presence of molecules are more important and could
be detected experimentally for weak resonances

We  start by reviewing  the many body theory for a condensate on a
Feshbach resonance. In a two component picture \cite{TTHK98} the mean field
hamiltonian density describing the trapped system is

\bea {\cal H}\!&=&\!\phi_a^*\left[-\frac{\hbar^2\nabla^2}{2m}
+V_a+\frac{\lambda_a}{2}\phi_a^*\phi_a\right]\phi_a \nonumber
\\
&+&\!\phi_m^*\left[-\frac{\hbar^2\nabla^2}{4m}  
 +V_m+\varepsilon+\frac{\lambda_m}{2}\phi_m^*\phi_m\right]\phi_m
 \nonumber\\
&+&\!\lambda_{am} \phi_a^*\phi_a\phi_m^*\phi_m +\frac{\alpha}{\sqrt{2}}
\Bigl[\phi^*_m\phi_a\phi_a+\phi_m\phi^*_a\phi^*_a\Bigr] ,\; \eea

\noindent where $\phi_{a(m)}$ denote the macroscopic wave function
of the atom (molecule) condensate, $\varepsilon$ is the detuning
parameter, $\alpha$ is the Feshbach resonance parameter introduced
in Ref.\cite{TTHK98}, and we assume that the confining potential
for the atoms and molecules are harmonic with axial symmetry,
$V_a=\frac{ m \omega_{a}^2}{2}(\beta^{a}_r
x^2+\beta^{a}_{r}y^2+\beta^{a}_{z}z^2)$,
$V_m=m \omega_{m}^2(\beta^{m}_r
x^2+\beta^{m}_{r}y^2+\beta^{m}_{z}z^2)$. As usual, the
Gross-Pitaevskii equations of motion for this model can be derived
by imposing that the action
\begin{equation}
\Gamma=\int dt \left\{ \;
d^3r \;i[\phi_a^*\dot\phi_a
+\phi_m^*\dot\phi_m
]
- E [\phi_a,\phi_m ] \right\} \;,
\label{action}
\end{equation}
\noindent where $E[\psi_a,\psi_m]=\int d ^3r \; {\cal H}\;$, is
stationary with respect to the variation of of $\phi_{a(m)}$,
subject to the normalization condition constraint
$\int d^3r(|\phi_a|^2+2|\phi_m|^2)=N.$
By looking for solutions of the form $\phi_a (r,t)=e^{-i\frac{\mu
t}{\hbar}}\phi_a (r)$ and  $\phi_m (r,t)=e^{-i\frac{2 \mu
t}{\hbar}}\phi_m (r)$ we find the coupled time-independent GP
equations,
\begin{eqnarray}
\mu\phi_a&=&\Bigl(-\frac{\hbar^2}{2m}\nabla^2+V_{a}
+\lambda_{a}|\phi_a|^2+\lambda_{am}|\phi_m|^2\Bigr)\phi_a \nonumber \\
&& +\sqrt{2}\alpha\phi_a^{\ast}\phi_m \label{phia} \label{atomic} \; \\
2\mu\phi_m&=&\Bigl(-\frac{\hbar^2}{4m}\nabla^2+\varepsilon+V_{m}
+\lambda_{am}|\phi_a|^2+\lambda_{m}|\phi_m|^2\Bigr)\phi_m \nonumber \\
&& +\frac{\alpha}{\sqrt{2}} \phi_a^{2}. \label{e2}
\end{eqnarray}
\noindent Now we turn our attention to the limit where the
two-component theory reduces to an effective one-component theory.
This limit is reached when
the detuning is large enough that $\varepsilon$ greatly exceeds the 
kinetic energy as well as any interaction energies, i.e., $\varepsilon>>
\lambda_{m} n_{m},\lambda_{am} n_{a}, \lambda_a n_{a}$, where $n_a$ and
$n_m$ are the atomic and molecular densities respectively,plus the
additional hypothesis that the number of molecules are small compared
to the number of atoms  
$n_{m}/n_{a} <<1$. These assumptions allow us to neglect the 
interaction terms and the chemical potential ($\mu \approx \lambda_a n_{a}$)
in the Gross-Pitaevskii equation (\ref{e2}), leading to a simple relation
between the molecular and atomic condensate
\begin{equation}
\phi_{m} \approx -\frac{\alpha}{\sqrt{2} \varepsilon} \phi_{a}^{2}\;.
\label{bobo}
\end{equation}
\noindent The insertion of this expression into equation (\ref{atomic}) 
yields an effective single condensate Gross-Pitaevskii equation 
for $\phi_{a}$
\begin{equation}
\mu \phi_{a} =  \left[- \frac{\hbar^{2} \nabla^{2}}{2 m} +V_{a}+
\lambda_{eff} \phi_{a}^{2} \right] \phi_{a}
\label{bobo1}
\end{equation}
\noindent with
\begin{equation}
\lambda_{eff}=\lambda_a-\frac{\alpha^2}{\varepsilon} \;.
\label{eff} 
\end{equation}
\noindent{Note} that,
from (\ref{bobo}), the limit $|\phi_{m}|/|\phi_{a}| <<1$ implies
$\varepsilon >> \alpha \phi_{a}$. As $\lambda_{a}$ is positive and we
want to vary $\varepsilon$ in a way that $\lambda_{eff}$ is always
positive, it follows from (\ref{eff}) that we must set $\varepsilon >
\alpha^{2}/\lambda_{a}$.Assuming that all the interaction strengths
are of the same order of magnitude, the conditions of validity of the 
one-component theory can be summarized by the inequalities:
$\varepsilon >>\lambda_{a}n_{a}$,$\varepsilon >> \alpha\sqrt{n_{a}}$
with $\varepsilon > \alpha^{2}/\lambda_{a}$. 
\noindent In what follows we will consider two regimes characterized
by the parameter r,
\begin{equation}
r = \frac{\alpha}{\lambda_{a} \sqrt{n_{a}}} . 
\label{A}
\end{equation}
\noindent which relates the characteristic energies of our system,
\begin{equation}
\alpha^{2}/\lambda_{a}= r \alpha\sqrt{n_{a}},\alpha\sqrt{n_{a}}= r
\lambda_{a}n_{a}
\label{ce}
\end{equation}
\noindent i)Strong resonance limit-When $r>>1 $ the characteristics 
energies satisfy the set of inequalities,$\alpha^{2}/\lambda_{a}>>
\alpha\sqrt{n_{a}}>>\lambda_{a}n_{a}$ and for the entire range of variation
of the detuning,even up to $\lambda_{eff}\approx 0 $, the conditions
of validity of the effective one-component theory is satisfied.
Therefore, when the parameter $r$ satisfies the above inequality
 one can picture the system as being one condensate of
atoms with an effective scattering length. We will show, in our
calculations that with values of $r$ in this range, the two
component and the effective one component theory agree even when
the effective scattering length is almost zero, the molecules are
practically inexistent in this regime. 
%

ii) Weak resonance limit -On the other hand when  $r \approx 1 $ it
follows that all the characteristics energies are of the same order 
of magnitude $\alpha^{2}/\lambda_{a} \approx \alpha\sqrt{n_{a}}
\approx \lambda_{a}n_{a} $ and for $ \varepsilon \approx \alpha^{2}
/\lambda_{a}$ ,the conditions of validity of the effective 
one-component theory is not satisfied.This is corraborated by our 
calculations where we show that, for detunings of the order of 
$\alpha^{2}/\lambda_{a} $ that correspond, in the single condensate effective
theory, to almost zero effective scattering length,  we  obtain significantly 
different equilibrium configuration and also have unique modes to the
hybrid system that are
relatively low in energy ($\omega_{exc} \sim \omega_{trap}$).

The weak resonance regime proves to be an ideal place to
experimentally detect a clear
signature of molecule formation. The expression for the parameter
$r$ illustrates that  there are three ways of approaching this
regime: by increasing the condensate density, having systems with a
higher asymptotic atom-atom scattering length and/or reducing the
value of $\alpha$.

Next we will show, as pointed out above, that the equilibrium spatial 
distribution of the atoms and the excitation  energies of specific
collective modes have different behaviors as a function of the
detuning $\varepsilon$ depending on the regimes discussed above and
can be a signature of the presence of molecules in the system.

The standard approach to investigate the properties of a hybrid
system would be to solve the equations (\ref{phia}), (\ref{e2})
and (\ref{bobo1}) to find the equilibrium shapes and to solve the linear
approximation of its time dependent extension to find the
collective excitations. However in this paper we use a simpler
approach, the variational method. In this method we employ a
Gaussian ansatz \cite{PG96,Lin02} to parameterize the time dependence of
the condensate wave function as
\begin{equation}
\phi_{a(m)}({\bf x},t) = A_{a(m)}({\bf x},t) e^{i F_{a(m)}({\bf x},t)}
\label{condensatepa}
\end{equation}
\noindent where the amplitudes of the condensate wave functions are
given by 
\bea A_{a} &=& \sqrt{\frac{N_{a}}{\pi^{3/2}}} \prod_{j=1,2,3}
\frac{1}{\sqrt{ q_{j}}}e^{-\frac{(x_{j}-r_{j}^{c})^{2}}{2 q^{2}_{j}}} \\
A_{m} &=& \sqrt{\frac{N_{m}}{\pi^{3/2}}} \prod_{j=1,2,3}
\frac{1}{\sqrt{Q_{j}}}
e^{-(\frac{(x_{j}-R_{j}^{c})^{2}}{2 Q^{2}_{j}}}, 
\eea
\noindent and the phase as
\bea
F_{a} \!&=&\! \theta_{a} + \sum_{j=1,2,3} \pi_{j} (x_{j}-r_{j}^{c}) +
\frac{p_{j}}{2 q_{j}} ( x_{j}- r_{j}^{c})^{2} \\
F_{m} \!&=&\! \theta_{m} +\! \sum_{j=1,2,3} \Pi_{j} (x_{j}-R_{j}^{c}) +
\frac{P_{j}}{2 Q_{j}} ( x_{j}- R_{j}^{c})^{2}.  \label{Fm}
\eea
\noindent The variational parameters $q_j$ $(Q_j)$, $r^{c}_{j}$
$(R^{c}_{j})$ correspond  respectively to the widths and
translation of the atomic and molecular condensate clouds, which
have an ellipsoidal shape $p_j$$(P_j)$, $\pi^{c}_{j}$
$(\Pi^{c}_{j})$ are the corresponding momenta. The $N_{a(m)}$ are the
number of particles of each component $\theta_{a(m)}$ the
corresponding phase.

When we replace the condensate wave function parametrized as in
eqs.(\ref{condensatepa}-\ref{Fm}) into equation (\ref{action}), 
$\Gamma$ reduces to a classical action
\bea 
\Gamma &=&\! \int \! dt \Bigl\{ 
\sum_{j=1,2,3} N_a \pi^c_j\dot r^c_j
+\frac{N_a}{4}\bigl(p_j\dot q_j - \dot p_j q_j\bigr)
\Bigr. \nonumber \\
&& \Bigl. +\sum_{j=1,\;2,\;3}
N_m \Pi^c_j\dot R^c_j +\frac{N_m}{4}\bigl(P_{j}\dot Q_{j}-\dot
P_{j}Q_{j}\bigr)
\Bigr. \nonumber \\
&& + \;\;\theta_a\dot N_a\;+\;
\theta_m\dot N_m \!
-E\Bigr\}
\eea
\noindent where the energy $E$ has the form $E=E_a+E_m+E_i+E_{_{FR}}$,
with $E_{a(m)}$ being the usual Gaussian
Gross-Pitaevskii energy for a single atomic(molecular) condensate
\cite{PG96}, $E_{i}$ are the interaction energies between these two
species and $E_{FR}$ is the Feshbach component.

The variation of the action with respect to the twenty eight
parameters leads to hamiltonian type equations of motion for
the time evolution of these parameters. Denoting collectively the
parameters by $X_{i}$, $i=1,\cdots,28$, the equations of motion
can be written as \cite{Lin02}
\begin{equation}
\sum_{l} \Gamma_{kl}({\bf X}) \dot{X}_{l} = \frac{\partial E}{\partial
X_{k}} ({\bf X}),
\label{eqmotion}
\end{equation}
\noindent where $\Gamma_{kl}({\bf X})$ is the anti-symmetric matrix
coming from the variation of the ``kinematic''piece of the action (14)

\noindent $E({\bf X})$ depends only on the phase difference,
$\theta= \theta_{a} - \theta_{m}/2$ and as a consequence the
quantity $N = N_{a} + 2 N_{m}$ is a constant of the motion. This
reduces the number of degrees of freedom to thirteen which we take
as the relative phase $\theta=\theta_{a} - \theta_{m}/2$ and the
relative population $n = (N_{a} - 2 N_{m})/2$, besides the
translational and shape degrees of freedom.

\medskip

i) \underbar{Equilibrium configuration}
\smallskip

The equilibrium configuration is determined by the condition
$\dot{{\bf X}_{k}}=0$, where $k=1,...,26$  which leads to
\be
\frac{\partial E}{\partial X_{k}} ({\bf X}_{0})=0\;,
\label{equations}
\ee
\noindent where ${\bf X}_{0}$ represents the equilibrium values of
the parameters. Thirteen of the equations (\ref{equations}) lead
immediately to $p_{j}=0 (P_{j}=0)$, ${\pi_{j}}^{c}=0
(\Pi_{j}^{c})=0$ and $\theta=\pi/2$ which is the phase of the minimum 
configuration. Six of the remaining
thirteen equations refers to the equilibrium displacement of the
condensate clouds. These equations have the trivial solution
$r^c_j=0 (R^c_j=0), j=1,2,3$, which is the one considered in this
paper. Therefore we are left with seven equations to find the
equilibrium widths of the condensate clouds and of the relative
population $n$, $n=(N_a-2N_m)/2$.

We proceed by presenting the results of the static calculations obtained
by solving (\ref{equations}) in the weak and strong resonance 
limits  respectively. In the strong 
resonance limit we used $r \approx 10$. We computed the equilibrium
radial and axial widths ($q_{x}$, $q_{y}$, $q_{z}$) for the atomic 
condensate for different 
values of the detuning. We varied the detuning from $2\pi\times$105 MHz to 
$2\pi\times$22 MHz corresponding, in the effective theory, to a
$\lambda_{eff}$ varying
from $80\%$ to $5\%$ of the asymptotic interaction strength,
respectively. The
results are shown on Table I.

One can see  no difference between the 
two component molecular theory and the single condensate  effective
theory. This shows that all attempts, through an adiabatic change of the
detuning, to  detect molecular signatures in this regime will fail,
because for all practical purposes the number of molecules formed is
quite insignificant. To achieve the weak resonant limit we used $r
\approx 1$. The detuning values were changed from $2\pi\times$41.7 KHz 
to $2\pi\times$8.7 KHz
corresponding to the same effective interaction strength  as in
the strong resonance case (from $80\%$ to $5\%$ of the asymptotic interaction
strength). The results, shown on Table II, indicate a very different 
behavior between the molecular and the single condensate effective 
theories, when the detuning approaches $ \alpha^{2}/\lambda_{a}$. 

In the two component theory, the size of the atomic condensate will not change
significantly when we vary the detuning, due to
the presence of molecules in contrast  with a drastic change in the
single condensate effective picture. We conclude that this regime 
proofs to be ideal for probing molecular signatures provided the
detuning is adiabatically changed. The weak resonance limit results are  
displayed in Fig.1. 

\begin{figure}
\centering
\begin{minipage}[c]{.45\textwidth}
\centering
\caption{Plot showing the atomic condensate width as a 
function of the detuning in the weak resonance limit ($r \approx 1$)
for (a) the radial width $q_x=q_y=q_{r}$ (b) the axial width
$q_{z}$. The dashed lined corresponds to the single component
effective theory and the continuous line the two component molecular
theory. The values of parameters used are:
$\omega_a=\omega_m=2\pi\times 500{\rm Hz}$, $\beta_r=1$,
$\beta_z=\frac{1}{15}$, $\lambda_a=\lambda_m=\lambda_{am}=120\mu{\rm
m}^3{\rm Hz}$, $\alpha=10^3\mu{\rm m}^{3/2}{\rm Hz}$, $N=10^5$ and
$m=23 u$}.   
\label{fig1}
\end{minipage}%
\hfill
\begin{minipage}[c]{.45\textwidth}
\begin{center}
\includegraphics[width=3.0in]{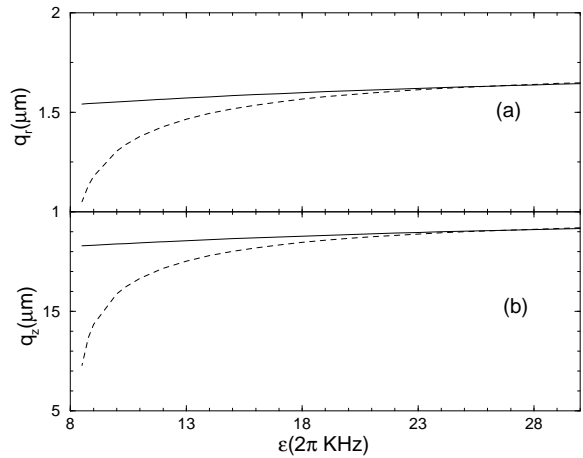}
\end{center}
\end{minipage}
\end{figure}

\medskip

\indent ii) \underbar{Excitations}

\smallskip

The excitations are associated with the normal modes of the linear
approximation of the equations of motion (\ref{eqmotion}). In this
limit (\ref{eqmotion})reduces to \cite{Lin02}
\begin{equation}
\sum_{l}\Gamma_{kl}({\bf X}_0)\dot{\eta_{l}} =
\sum_{l}{ {\cal H}}_{kl}({\bf X}_0)\eta_{l}\;.
\label{soeq}
\end{equation}
\noindent{where} ${\cal H}_{kl}({\bf X}_0)$ is the Hessian matrix
at equilibrium ${\rm H}_{kl}=\frac{\partial^2 E }{\partial{
X}_k\partial{X}_l}( {\bf X}^0)$ and $\eta_l$ are the displacement
from the equilibrium value $X_l=X_{l0}+\eta_l$. The normal modes
correspond to oscillatory solutions of (\ref{soeq}), and the
eigenfrequencies are given
by the characteristic equation ${\rm det}(i \Gamma \Omega-{\cal H})=0$. 
The roots of the caracteristic equation  are real and come out in 
pairs $(\Omega,-\Omega)$, each pair associated to one normal mode.

In the single-condensate case we have six normal modes which are related to
the translations of the condensate clouds and to shape
oscillations \cite{PG96}. For the later, two of these correspond to axially symmetric
oscillations where the axial and radial oscillation are in phase
(mode c) and the other where they are out of phase (mode b), the
third one being non-axially symmetric with the oscillations in $x$
and $y$ directions out of phase and the axial direction at its
equilibrium value (mode a).

In the two component case the normal modes will be linear
combinations of the modes described above for each component plus
one mode related to the dynamics of population exchange between
atoms and molecules, giving a total of thirteen normal modes. In what
follows we will investigate how the eigenfrequencies depend on the
Feshbach resonance parameter $\alpha$ and the detuning $\varepsilon$.

In our calculation we obtain the following general results, independent
of the value of $r$: a)The six translational degrees of freedom are 
decoupled from the shape and number degrees of freedom, originating three
atomic and molecular translations in phase (center of mass modes) and
three out of phase (dipole modes).  
b)The remaining seven modes can be divided into two groups. The first group,
where the shape and number oscillations are decoupled, consists of two 
modes which correspond to an in phase and an out of phase atomic and
molecular oscillations where both components are in  mode a.  The second 
group consists of coupled shape and number oscillations with each
component in  modes b and/or c.      
c) In the single component theory we have six modes, three
translational and three shape oscillations. The three translational
modes correspond to the three center of mass modes of the hybrid
system and the three shape oscillations correspond to the three lowest
energy modes not counting the center of mass modes. 
 
For the strong resonance limit, $r \approx 10$, the three
lowest energy modes, disrecarding the translational ones, for the
hybrid system correspond to  
in phase atomic and molecular shape oscillations (b+b), (a+a) and (c+c)
respectively, and have the same behavior in both theories as
exemplified on Table I for the mode (a+a).

In the weak resonance limit $r \approx 1$ of the three modes only the
mode (a+a) behaves differently  depending on which theory is used. In
the single component theory the a mode approaches abruptly to the c
mode when $\lambda_{eff} \rightarrow 0$, because of the degenerescence
present in the ideal gas limit. On the other hand for the hybrid
theory this abrupt change is not present. The absence of a near 
degenerescence is a signature of the presence of the molecules. The
values of this frequency are shown on Table II and two of the three
lowest energy modes, for both single and hybrid theories are displayed
in Fig.2. 

\begin{figure}
\centering
\begin{minipage}[c]{.45\textwidth}
\centering
\caption{
Plot comparing two of the mode energies, in units of the radial 
trap frequency, as a function of the detuning. The dashed lines
correspond to the single component effective theory and the continuous line
the two component molecular theory. The lowest mode and four highest
modes are not shown in the figure. See caption of figure 1 for the
parameters. }
\label{fig2}
\end{minipage}%
\hfill
\begin{minipage}[c]{.45\textwidth}
\begin{center}
\includegraphics[width=3.0in]{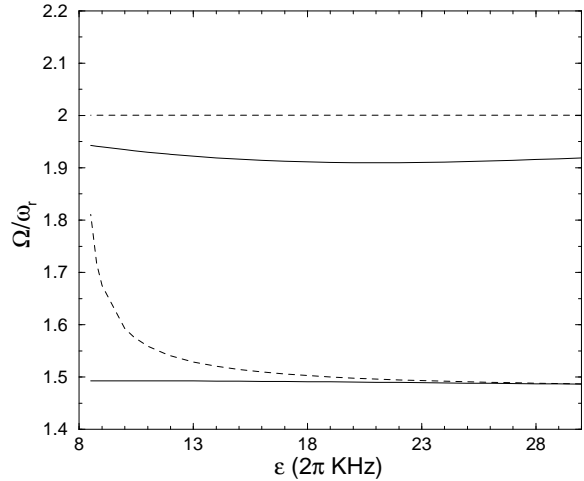}
\end{center}
\end{minipage} 
\end{figure}

Another possible signature of a molecular condensate is the presence
of the dipole mode frequencies in the two componet theory.   The  
frequency of the center of mass oscillation motion of the hybrid 
BEC is entirely determined by the
frequencies of the trap and does not depend on the interatomic
interaction \cite{DGPS99}. On the other hand the frequencies of the dipole 
oscillation of the hybrid system depend on the Feshbach
resonance interaction parameter, $\alpha$, and on the detuning
parameter $\varepsilon$. Since this mode is unique to the hybrid system
its presence is a clear signature of the presence of molecules. This
being an interesting  phenomenon to be explored experimentally. Of the
three dipole modes the radial ones are degenerate and have a higher
frequency than the axial one. Using the same parameters for the weak
limit we plot in Fig.3 the frequency of this mode as a function of the
detuning. From this results we see that the frequency of this mode,
unique to the hybrid system, decreases when the detuning approaches
$\alpha^{2}/\lambda_{a}$ reaching observable values.

\begin{figure}
\centering
\begin{minipage}[c]{.45\textwidth}
\centering
\caption{
Plot showing the  dipole frequency, in units of the radial 
trap frequency, as a function of the detuning for the
hybrid  condensate. The lower curve corresponds to an axial 
oscillation  and the upper curve radial oscillations.  The parameters
are the same as in figure 1. See text for details
}
\label{fig3}
\end{minipage}%
\hfill
\begin{minipage}[c]{.45\textwidth}
\begin{center}
\includegraphics[width=3.0in]{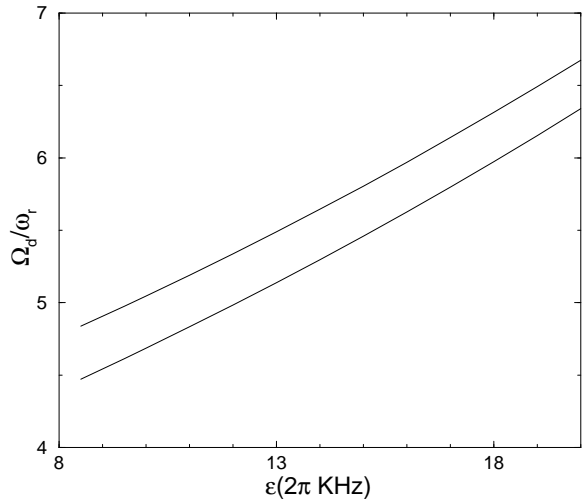}
\end{center}
\end{minipage}
\end{figure}

In conclusion, we have shown, that provided we work in the weak
resonance limit, the hybrid atom molecule BEC system behaves
qualitatively differently  from a single condensate with an
effective scattering length, exhibiting  different behaviours for the
equilibrium shapes and collective excitations as a function of the
detuning. Besides for reasonable values of the physical parameters 
it appears a unique mode of the hybrid system whose frequencies are 
found to be of the order 
of the trap frequencies for detuning 
values of the order of KHz making the observation of this
mode experimentally feasible.

\setcounter{table}{0}
\begin{table}
\label{tbl}
\caption{Comparison between the single condensate theory and the hybrid
theory in the strong resonant limit. The Feshbach resonance parameter
in this case is $\alpha=5 10^4\mu{\rm m}^{3/2}{\rm Hz}$} and the other
parameters are the same as in figure 1.
\centering
\vspace{.2cm}
\begin{tabular}{|r|r|r|r|r|r|r|r|} \hline
\multicolumn{8}{|c|}{Strong Resonance Limit} \\ \hline
\multicolumn{1}{|c|}{} & \multicolumn{4}{c|}{Single Condensate Effective Theory} &
\multicolumn{3}{|c|}{Molecular  Theory} \\ \hline
$\varepsilon (2\pi\times {\rm MHz})$ & $\lambda/\lambda_{eff}$ & 
$q_{r}(\mu m)$ & $q_{z}(\mu m)$ & $\omega_{a}/\omega_{r}$ &
$q_{x}(\mu m)$ & $q_{z}(\mu m)$ & $\omega_{a}/\omega_{r}$ \\ \hline
 105  & 0.8 & 1.7 & 23.9 & 1.5 & 1.7 & 23.9 & 1.5 \\ \hline
 30.0 & 0.3 & 1.4 & 19.3 & 1.5 & 1.4 & 19.3 & 1.5 \\ \hline
 26.0 & 0.2 & 1.3 & 17.5 & 1.6 & 1.3 & 17.5 & 1.6 \\ \hline
 22.0 & 0.1 & 1.1 & 12.6 & 1.7 & 1.2 & 12.8 & 1.7 \\ \hline
\end{tabular}
\end{table}

\begin{table}
\label{tb2}
\caption{Comparison between the single condensate effective theory and the hybrid
molecular theory in the weak resonant limit.}
\centering
\vspace{.2cm}
\begin{tabular}{|r|r|r|r|r|r|r|r|} \hline
\multicolumn{8}{|c|}{Weak Resonance Limit} \\ \hline
\multicolumn{1}{|c|}{} & \multicolumn{4}{c|}{Single Condensate Effective Theory} &
\multicolumn{3}{|c|}{Molecular  Theory} \\ \hline
$\varepsilon (2 \pi {\rm KHz})$ & $\lambda/\lambda_{eff}$ & $q_{r} (\mu m)$ 
& $q_{z}(\mu m )$ & $\omega_{2}/\omega_{r}$ 
& $q_{r}(\mu m)$ & $q_{z} (\mu m)$ & $\omega_{a}/\omega_{r}$ \\ \hline
 41.7 & 0.8 & 1.7 & 23.9 & 1.5 & 1.7 & 23.8 & 1.5 \\ \hline
 11.9 & 0.3 & 1.4 & 19.3 & 1.5 & 1.6 & 22.0 & 1.5 \\ \hline
 10.4 & 0.2 & 1.3 & 17.5 & 1.6 & 1.6 & 21.8 & 1.5 \\ \hline
 8.7  & 0.1 & 1.1 & 12.6 & 1.7 & 1.5 & 21.6 & 1.5 \\ \hline
\end{tabular}
\end{table}

\smallskip

{\bf ACKNOWLEDGMENT}: This work was partially supported by
Fundac\c{c}\~ao de Amparo $\grave{a}$ Pesquisa do Estado de
S$\tilde{a}$o Paulo (FAPESP) under contract number 00/06649-9. E.
J. V. de Passos and M.S. Hussein are supported in part by CNPq.
The work of Chi-Yong Lin was supported in part by
the National Science Council, ROC under the Grant
NSC-89-2112-M-259-008-Y.


\medskip

\begin{thebibliography}{99}
%
\bibitem{An95}
M.H. Anderson, J.R. Ensher,
M.R. matthews, C.E. Wieman, and E.A. Cornem, {\it Science} 
{\bf 269}, 198 (1995); K.B. Davis, M.-O. Mewes, M.R. Andrews, N.J. van
Druten, D. S. durfee, D.M. Kurn, and W. Ketterle, {\it
Phys. Rev. Lett} {\bf 75}, 3969(1995).
%
\bibitem{DGPS99}
See, e.g., F. Dalfovo, S. Giorgini, L.P. Pitaevskii and S.
Stringari, Rev. Mod. Phys. {\bf 71}, 463 (1999);
 A.J.Leggett, Rev. Mod. Phys. {\bf 71}, 463 (2001).
%
\bibitem{Ha98}
M. R. Matthews, D.S. Hall, D. S. Jin, J. R. Ensher  C.E. Wieman and
E.A. Cornell, 
Phys. Rev. Lett. {\bf 81}, 243 (1998);
D.S. Hall, M.R. Matthews, C.E. Wieman and E.A. Cornell,
Phys. Rev. Lett. {\bf 81}, 1539 (1998);
D.S. Hall, M.R. Matthews, C.E. Wieman and E.A. Cornell,
Phys. Rev. Lett. {\bf 81}, 1543 (1998).
%
\bibitem{In98}
S. Inouye, M. R. Andrews, J. Stenger, H.-J.Miesner, D. M.
Stamper-Kurn and W. Ketterle, Nature {\bf 392}, 151 (1998);
J. Stenger, S. Inouye, M. R. Andrews,  H.-J.Miesner, D. M.
Stamper-Kurn and W. Ketterle, Phys. Rev. Lett. {\bf 82}, 2422
(1999).
%
\bibitem{TTHK98}
E. Timmermans, P. Tommasini, R.
C\^ote, M. S. Hussein and A. K. Kerman, Phys. Rev.
Lett. {\bf 83}, 2691 (1999);
P. Tommasini, E. Timmermans, M. S. Hussein and A. K. Kerman, Phys.
Rep. {\bf 315}, 199 (1999).
%
\bibitem{Dr98}
P. D. Drummond, K. V.  Kheruntsyan, Phys. Rev. Lett. {\bf 81}, 
3055 (1998); M. Holland, Phys. Rev. Lett. {\bf 86}, 1915 (2000);
A. Vardi,  Phys. Rev. A {\bf 64}, 063611-1 (2001).
%
\bibitem{Wy00}
R. Wynar, R. S. Freeland, D. J. Han, C. Ryu, D. J. Heinzen, Science,
{\bf 287}, 1016 (2000);
%
\bibitem{HWDK00}
D. J. Heinzen, R. Wynar, P. D. Drummond and Kheruntsyan,
Phys. Rev. Lett. {\bf 84}, 5029 (2000).
%
\bibitem{Wieman00}
C. Wieman, Rb-85, through a Feshbach resonance (attractive and
repulsive $a$) ``Supernova'' effect of the condensate, APS News
(2002).
%
\bibitem{HO01}
J. J. Hope and M. K. Olsen, Phys. Rev. Lett. {\bf 86}, 3220 (2001).
%
\bibitem{Do02}
Elizabeth A. Doley, Neil R. Claussen, Sarah T. Thompson and Carl
E. Wieman, Nature {\bf 417},529 (2002).
%
\bibitem{PG96}
V. M. P\'erez-Garcia, H Michinel, J. I. Cirac, M. Lewenstein and
P. Zoller, Phys. Rev. Lett. {\bf 77}, 5320 (1996). 
%
\bibitem{Lin02}
C-Y. Lin, E.J.V. de Passos, M. S. Hussein, D-S. Lee and 
A.F.R. de Toledo Piza, Physica A {\bf 318 }, 423 ( 2003).
\end{thebibliography}
\end{document}